\documentclass[useAMS,usenatbib]{mn2e}
\usepackage{Times}
\usepackage{epsfig}

\newcommand{\CIV}{C\,{\sc iv}}
\newcommand{\MgII}{Mg\,{\sc ii}}
\newcommand{\FeII}{Fe\,{\sc ii}}

\newcommand{\Hbeta}{H{\sc$\beta$}}  

\def\mbh{{$\mathcal M_{\textrm{BH}}$}}

\def\msol{{$\mathcal M_{\odot}$}}

\def\kmps{{km~s${}^{-1}$}}
\def\ergps{{erg~s${}^{-1}$}}
\def\percm3{{cm${}^{-3}$}}

\def\apj{{ApJ, }}

\def\apjl{{ApJL, }}

\title[Biases in the Quasar Mass-Luminosity Plane]{Biases in the Quasar Mass-Luminosity Plane}
\author[A. Rafiee and P. B. Hall]{Alireza Rafiee and Patrick B. Hall
\\
Department of Physics and Astronomy, York University, 4700 Keele St., Toronto, Ontario, M3J 1P3, Canada\\}

\begin{document}
\date{Accepted 2011 April 13. Received 2011 March 30; in original form 2010 November 04}

\maketitle

\label{firstpage}

\begin{abstract}

We find that the recently reported departure from the Eddington luminosity limit
for the highest quasar black hole masses at a given redshift
is an artifact due to biases in black hole mass measurements.
This sub-Eddington boundary (with non-unity slope)
in the quasar mass-luminosity plane
was initially reported by Steinhardt \& Elvis (2010a)
using the FWHM-based black hole mass catalogue of Shen et al. (2008).
However, the significance of the boundary is reduced when the FWHM-based
mass-scaling relationship is recalibrated following Wang et al. (2009) and
using the most updated reverberation mapping estimates of black hole masses.
Furthermore, this boundary is not seen using mass estimates based on
the line dispersion of the same quasars' \MgII\ emission lines.
Thus, the initial report of a sub-Eddington boundary
with non-unity slope was due to biases in estimating masses using the FWHM of a fit
of one or two Gaussians to quasar \MgII\ emission lines.
We provide evidence that using the line
dispersion of the \MgII\ line produces less biased black hole mass estimates.

\end{abstract}


\section{Introduction}\label{Sec:Introduction_SEB}

Steinhardt \& Elvis (2010a) (hereafter SE10a; see also Steinhardt \& Elvis 2010bcd) have claimed that there exists a departure from the Eddington luminosity ($L_{Edd}$) boundary for the highest quasar black hole masses at a given redshift; in other words, a sub-Eddington boundary (SEB) with non-unity slope.
SE10a investigated the authenticity of this SEB and argued that its presence was neither due to measurement error nor due to small number statistics.

SE10a used the black hole (BH) masses of Shen et al. (2008), who adopted mass scaling relations for \Hbeta\ and \CIV\ from Vestergaard \& Peterson (2006) and for \MgII\ from McLure \& Dunlop (2004).
Both those references have made two key assumptions.
They have assumed a tight relationship between the radius of the emitting region and the AGN continuum luminosity in the form of $R \propto L^\beta$, where $\beta\simeq 0.5$.
They have also assumed a relationship between the characteristic virial velocity of the gas emitting a particular emission line, $v_{virial}$, and the single-epoch FWHM of the same line of the form $v_{virial} \propto $ FWHM.
SE10a argued that departures from the above assumptions sufficient to affect
the presence of a SEB were unlikely, although they acknowledged that the slope
of a SEB is dependent upon the BH mass scaling relation used (their \S 4.5.2).

Peterson et al. (2004) suggested that the BLR virial velocity can be characterized by either the line dispersion $\sigma_{line}$ or the FWHM measured directly from the emission line profile in an rms spectrum\footnote{In a reverberation mapping study, the mean spectrum is defined as $\overline{F(\lambda)}=\frac{1}{N}\sum_{i=1}^{N}F_{i}(\lambda)$ where $F_{i}(\lambda)$ is the \emph{i}th spectrum of the N spectra that compose the reverberation data for one object. The rms spectrum is defined as $S(\lambda)=\left\{\frac{1}{N-1}\sum_{i=1}^{N}[F_{i}(\lambda)-\overline{F(\lambda)}]^{2}\right\}^{1/2}$ (Peterson et al. 2004).} obtained during a reverberation mapping study (see, e.g., Figure 3 of Peterson et al. 2004). This result has been interpreted as arising from linear relations between virial velocity and each of the rms FWHM and rms $\sigma_{line}$. The same study shows that the line dispersion measured from a mean spectrum is also linearly proportional to the virial velocity; however, the FWHM measured from a mean spectrum may not be linearly proportional to the virial velocity (see e.g., Figure 4 \& 5 of Peterson et al. 2004). Since single-epoch spectra are more similar to a mean spectrum than an rms spectrum, Wang et al. (2009) claim that the relation $v_{virial} \propto $ FWHM may not hold in general when we use single epoch spectra and suggest using a relationship of the form $v^2_{virial} \propto \textrm{FWHM}^{\gamma}$ instead.  The virial relation still holds the same form of \mbh~$\propto R~v_{virial}^{2}$.

Motivated by Wang et al. (2009), we investigate the reliability and biases of the SEB by re-estimating the mass scaling relation and by using two different methods of estimating line width; namely, using the FWHM and the line dispersion of a single epoch spectrum.
We present and discuss the SEB in \S\,\ref{sec:Sub_Eddington_Boundary}.
We compare the use of FWHM- and line dispersion-based BH masses in \S\,\ref{sec:FWHMsigma}.
We discuss the effects of using different mass scaling relationships on the SEB in \S\,\ref{sec:BH_Mass_rescaling}.
We end with our conclusions in \S\,\ref{sec:Conclusion_SEB}.

\section{The Sub-Eddington Boundary}\label{sec:Sub_Eddington_Boundary}

SE10a have plotted the mass-luminosity plane using the black hole masses of Shen et al. (2008) from the SDSS DR5 quasar sample (Schneider et al. 2007). They have used 62185 quasars over the redshift range $0.2 < z < 4.0$ with three BH mass scaling relations using the FWHMs of the \Hbeta, \MgII, and \CIV\ emission lines.

SE10a divide their \Hbeta, \MgII, and \CIV\ mass estimates into 13 redshift bins (though their conclusions are only based on \Hbeta\ and \MgII). They show that SEB of non-unity slope is a prominent feature in most of the bins, especially the highest redshift ones. The \MgII\ emission line can be measured from SDSS spectra in eight of these redshift bins.

\begin{figure*}
 \includegraphics[width=3.0in,height=2.5in]{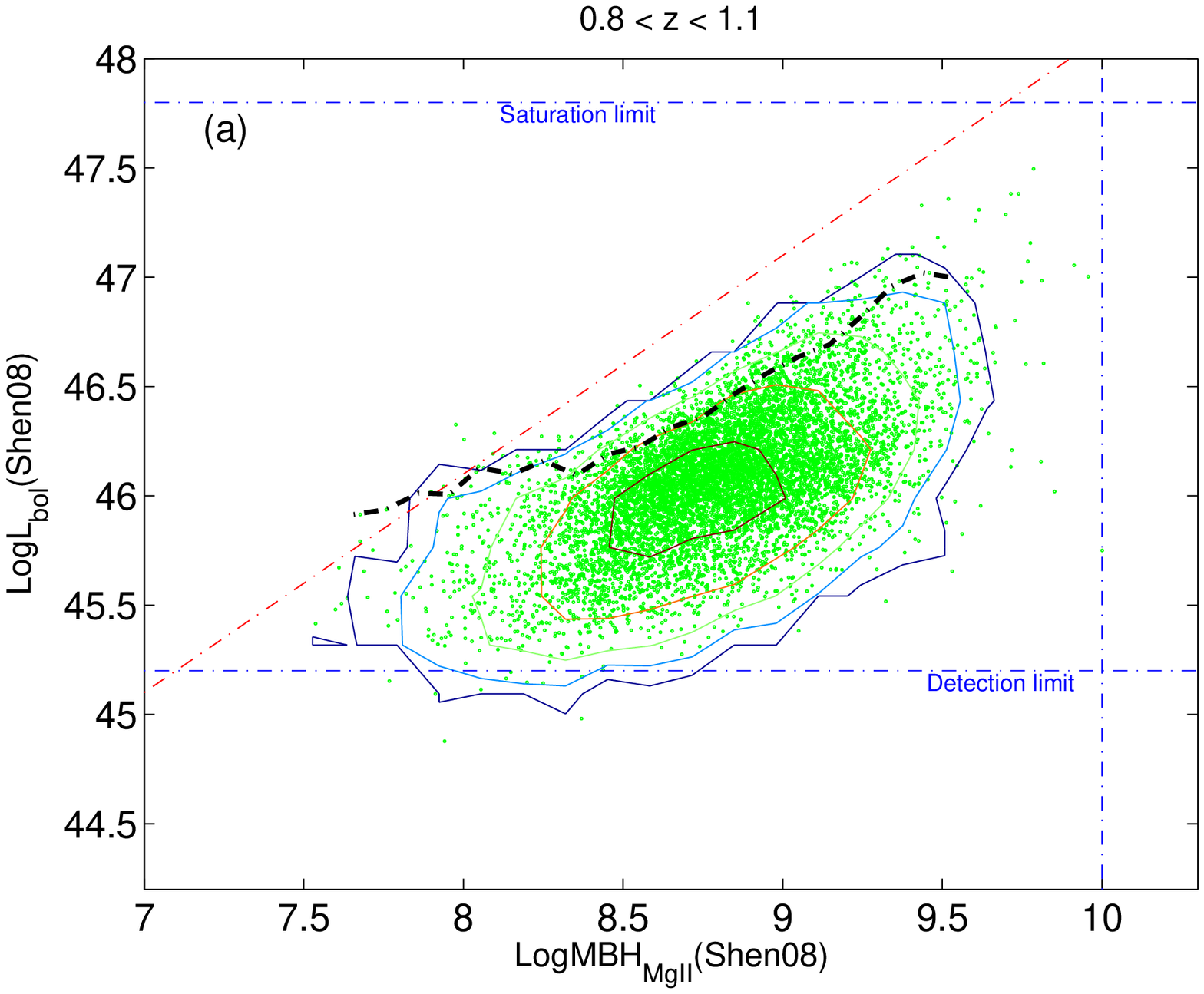}
 \includegraphics[width=3.0in,height=2.5in]{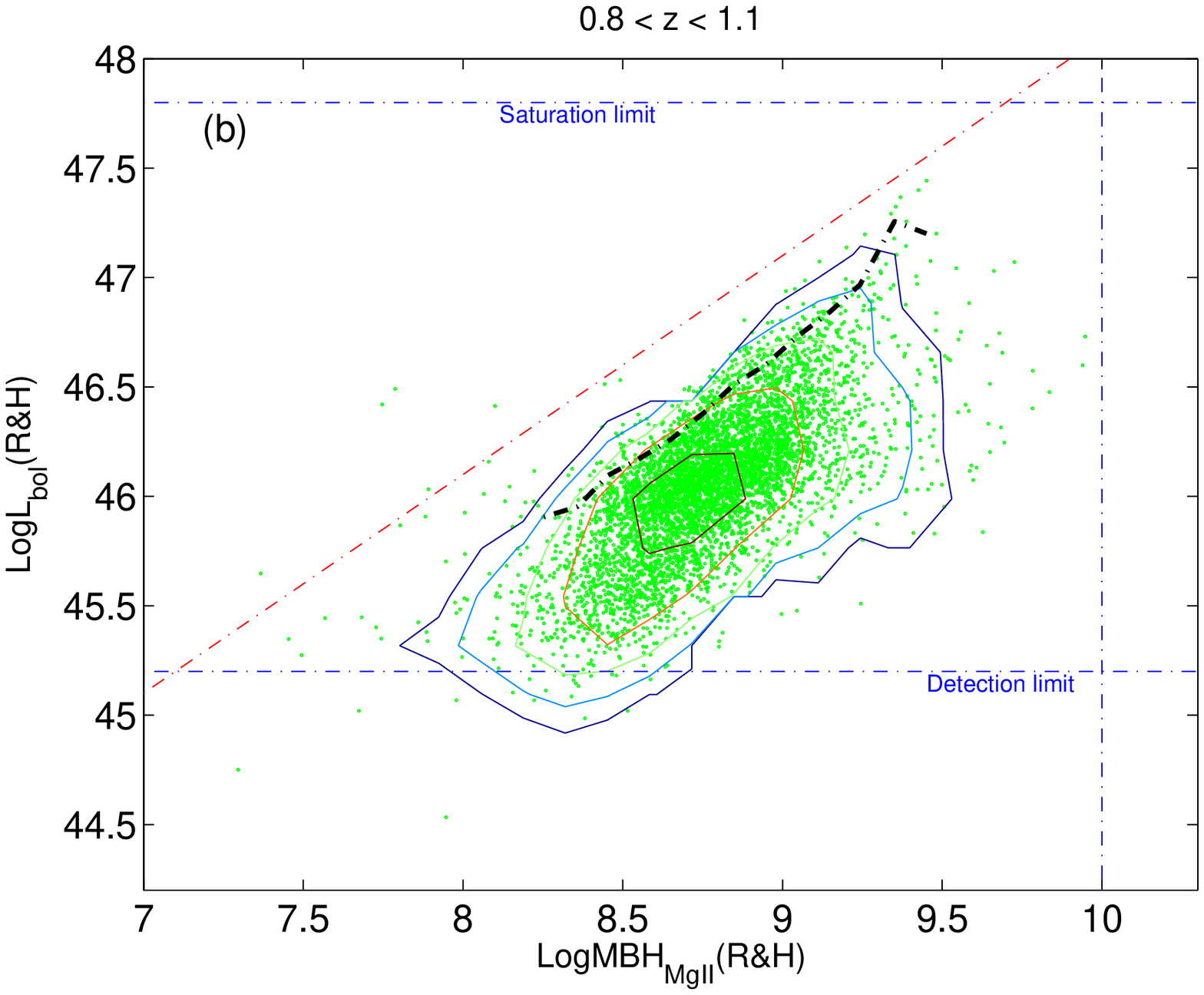}\\
  \caption{The quasar \MgII\ mass-luminosity plane, (a) using the original BH masses of Shen et al. (2008), and (b) using the BH masses of Rafiee \& Hall (2011). 
  The lower horizontal dot-dashed line is the approximate lower luminosity limit for a low-redshift object at the faint magnitude limit of the SDSS.
  The upper horizontal dot-dashed line is the upper luminosity limit for a high-redshift object at the bright saturation limit of the SDSS. Objects below the detection limit or above the saturation limit are not in the SDSS spectroscopic sample. The vertical dot-dashed line represents an upper mass limit of $\log$\mbh\ $=10$.
SE10a have claimed there exists a sub-Eddington boundary of non-unity slope, which is here illustrated in panel (a) for objects with
$\log$\mbh\ $> 8.2$ and $\log L_{bol} >46.1$ 
as the region between the black dot-dashed curve (the $95\%$ upper luminosity limit of the distribution as a function of mass) and the red dot-dashed line (an Eddington ratio of one).}\label{Fig:quasar_mass_luminosity_plane}
\end{figure*}

In Figure \ref{Fig:quasar_mass_luminosity_plane}a, we illustrate the SEB seen by SE10a using the Shen et al. (2008) quasar BH masses in a particular redshift bin. The gap between the red dot-dashed line (showing an Eddington ratio of one) and the black dot-dashed curve (tracing the 95th percentile of the observed distribution) is called the sub-Eddington boundary.

\begin{figure*}
  \includegraphics[width=5in,height=4in]{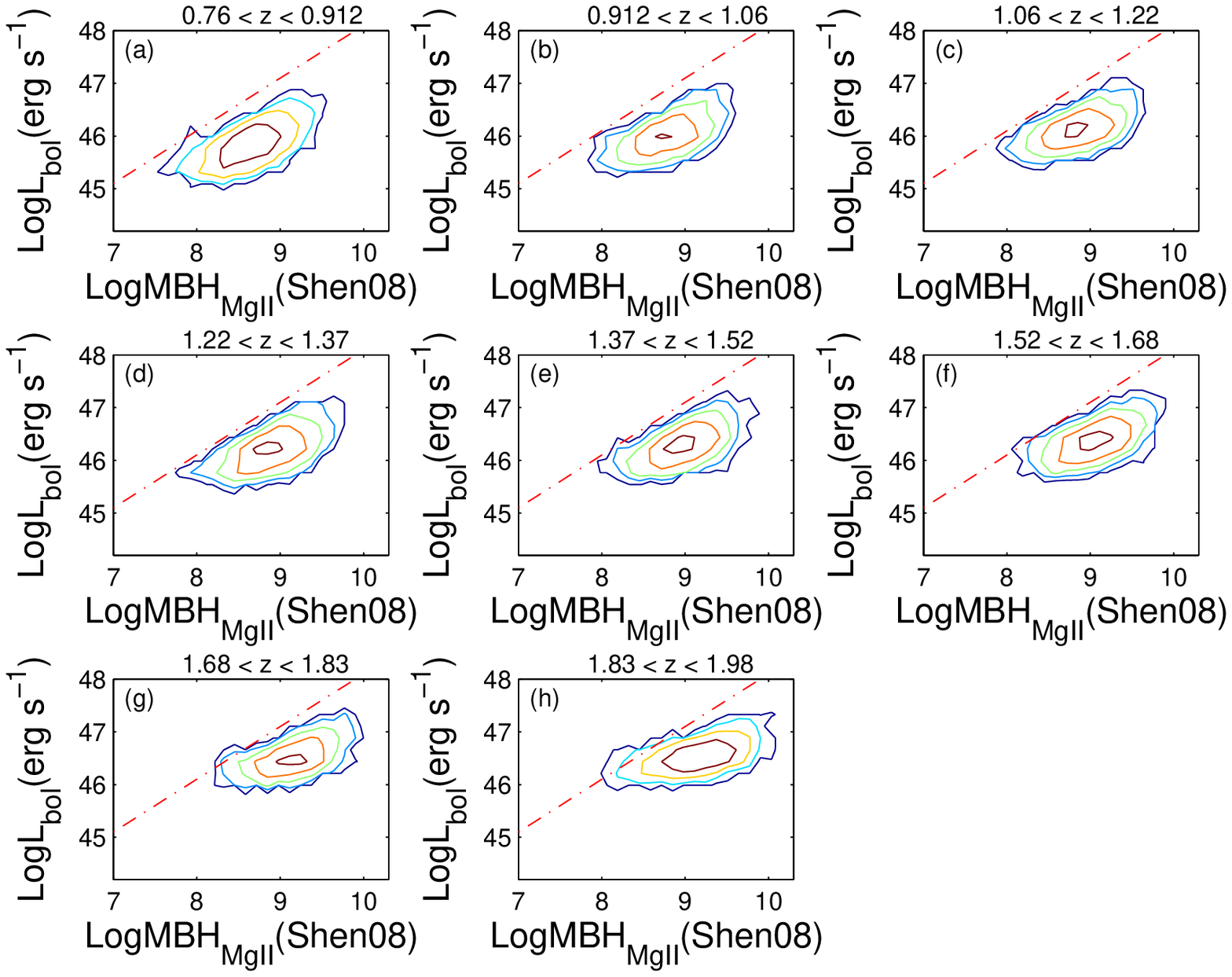}\\
 \caption{The quasar \MgII\ mass-luminosity plane for 8 redshift bins using the original BH masses from Shen et al. (2008). A sub-Eddington boundary of non-unity slope for objects with the highest \mbh\ in a given redshift bin can be seen, as found by SE10a.}\label{Fig:8quasar_mass_luminosity_plane_original}
\end{figure*}

In Figure \ref{Fig:8quasar_mass_luminosity_plane_original} panels (a) to (h), we use the same quasar sample as Figure \ref{Fig:quasar_mass_luminosity_plane}a but plot the mass-luminosity plane for 8 redshift bins between $0.76 < z < 1.98$. Based on a similar diagram (their Figure 8, panels 4 to 11), SE10a concluded that the most luminous low-mass quasars in any given redshift bin {\em are} at the Eddington limit but that the most luminous high-mass quasars are {\em not} at the Eddington limit.

\begin{figure*}
  \includegraphics[width=5in,height=4in]{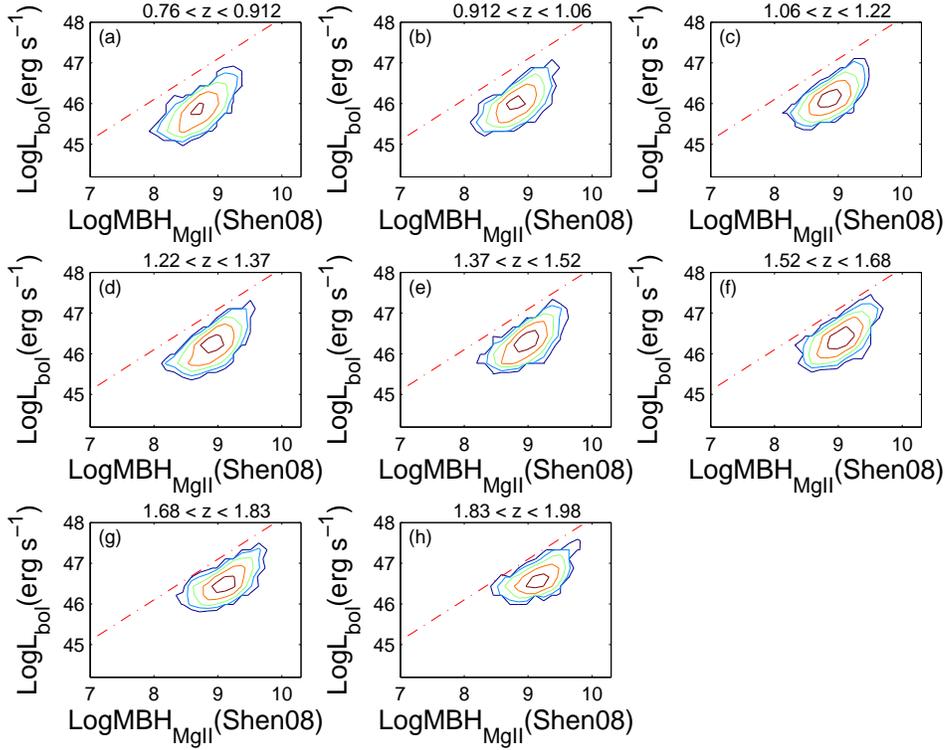}\\
 \caption{The quasar \MgII\ mass-luminosity plane for 8 redshift bins when we use the $\sigma_{line}$-based BH masses of the Rafiee and Hall (2011) catalogue.  The slope of the upper envelope of the quasar distribution is consistent with unity, unlike what was found by SE10a.}\label{Fig:8in1_zbin_Lbol_MBH_RH_calibration}
\end{figure*}

However, a SEB with non-unity slope cannot be seen using the Rafiee \& Hall (2011) BH mass catalogue which is based on the line dispersion of \MgII\ instead of its FWHM.  This catalogue consists of 27602 quasars with $0.7 < z < 2.0$, where the redshift limits are set by our method of removing the \FeII\ pseudocontinuum from the \MgII\ region of the spectrum.
In Figure \ref{Fig:quasar_mass_luminosity_plane}b and Figure \ref{Fig:8in1_zbin_Lbol_MBH_RH_calibration} panels (a) to (h), we plot the mass-luminosity plane for the same objects as in Figure \ref{Fig:8quasar_mass_luminosity_plane_original} but using the Rafiee \& Hall (2011) BH mass estimates.
This plot shows no evidence for the existence of an SEB with slope different from
unity.  Instead, in the lower redshift bins there is a sub-Eddington boundary whose slope is consistent with unity; i.e., a boundary at a fixed fraction of the Eddington limit.  As the redshift of the bin increases, the boundary moves to a higher fraction of the Eddington limit until it reaches the Eddington limit itself.

Comparing Figure \ref{Fig:quasar_mass_luminosity_plane}a with Figure \ref{Fig:quasar_mass_luminosity_plane}b or comparing Figure \ref{Fig:8quasar_mass_luminosity_plane_original} with Figure \ref{Fig:8in1_zbin_Lbol_MBH_RH_calibration} shows that two BH mass samples from the same SDSS data set yield different distributions in the mass-luminosity plane.
The fact that different mass-scaling relations were used for these samples suggests that the SEB is sensitive to the mass calibration or the line width indicator used, or both.
We now proceed to investigate these line width indicator and mass calibration
issues.

\section{Characterizing the Line Width}\label{sec:FWHMsigma}

Shen et al. (2008) fit \MgII\ as a sum of two Gaussians, one constrained to have FWHM$<$1200~\kmps, and use the FWHM of the broader Gaussian as the \MgII\ FWHM. It is convenient to assume that quasar emission lines can be modeled as sums of two Gaussians; however, the validity of this Gaussianity assumption has not yet been fully investigated. There are only a few studies on the non-Gaussianity of quasar broad emission lines in the context of BH mass estimation. Peterson et al. (2004) have suggested using the line dispersion and not the FWHM of the emission profile for mass measurements for several reasons including the Gaussianity assumption. Collin et al. (2006) have used both the FWHM and rms line dispersion of reverberation-mapped AGN to investigate the Gaussianity of the lines. They reported two classes of objects according to the value of ${\rm FWHM}/{\sigma_{line}}$: Pop.~I with values below the Gaussian ratio of $2\sqrt{2\ln 2}\simeq2.35$ and Pop.~II with values above it.  Peterson (2007) have also reported that for H$\beta$ this ratio spans $0.71 < {\rm FWHM}/{\sigma_{line}} < 3.45$ with an average ratio of $2.0$, indicating that the emission lines may not be well fit by single or double Gaussians.

\begin{figure*}
  \includegraphics[width=3in,height=2.5in]{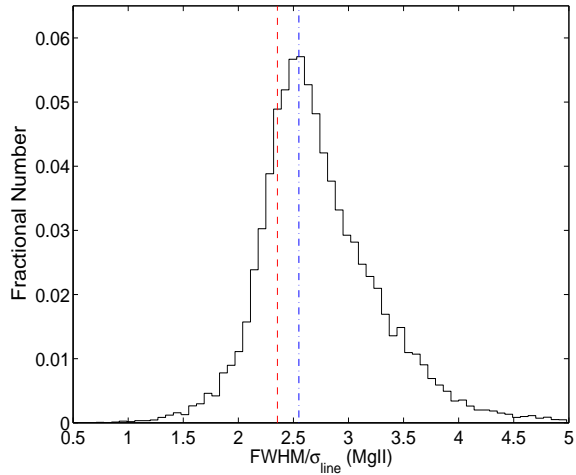}\\
 \caption{Histogram of $FWHM/\sigma_{line}$ ratios for the \MgII\ emission line. The FWHM is taken from Shen et al. (2008) and $\sigma_{line}$ is taken from Rafiee \& Hall (2011). The red dashed line represents the value of $2\sqrt{2\ln 2}$ corresponding to a perfect Gaussian profile. The blue dash-dotted line represents the peak at 2.55.}\label{Fig:histogram_FWHM2sigma_ratio}
\end{figure*}

We have investigated the significance of the Gaussianity assumption for BH mass estimates using the line dispersion $\sigma_{line}$ of all objects from Rafiee \& Hall (2011) and the FWHM reported by Shen et al. (2008) for the same objects.  In Figure \ref{Fig:histogram_FWHM2sigma_ratio}, we have plotted the histogram of the resulting \MgII\ ${\rm FWHM}/\sigma_{line}$ distribution.  We report a range of $1 < {\rm FWHM}/{\sigma_{line}} < 5$ with a mean value of $2.7$ and a mode of $2.55$ (the blue dash-dotted line in Figure \ref{Fig:histogram_FWHM2sigma_ratio}).

\begin{figure*}
  \includegraphics[width=3in,height=2.5in]{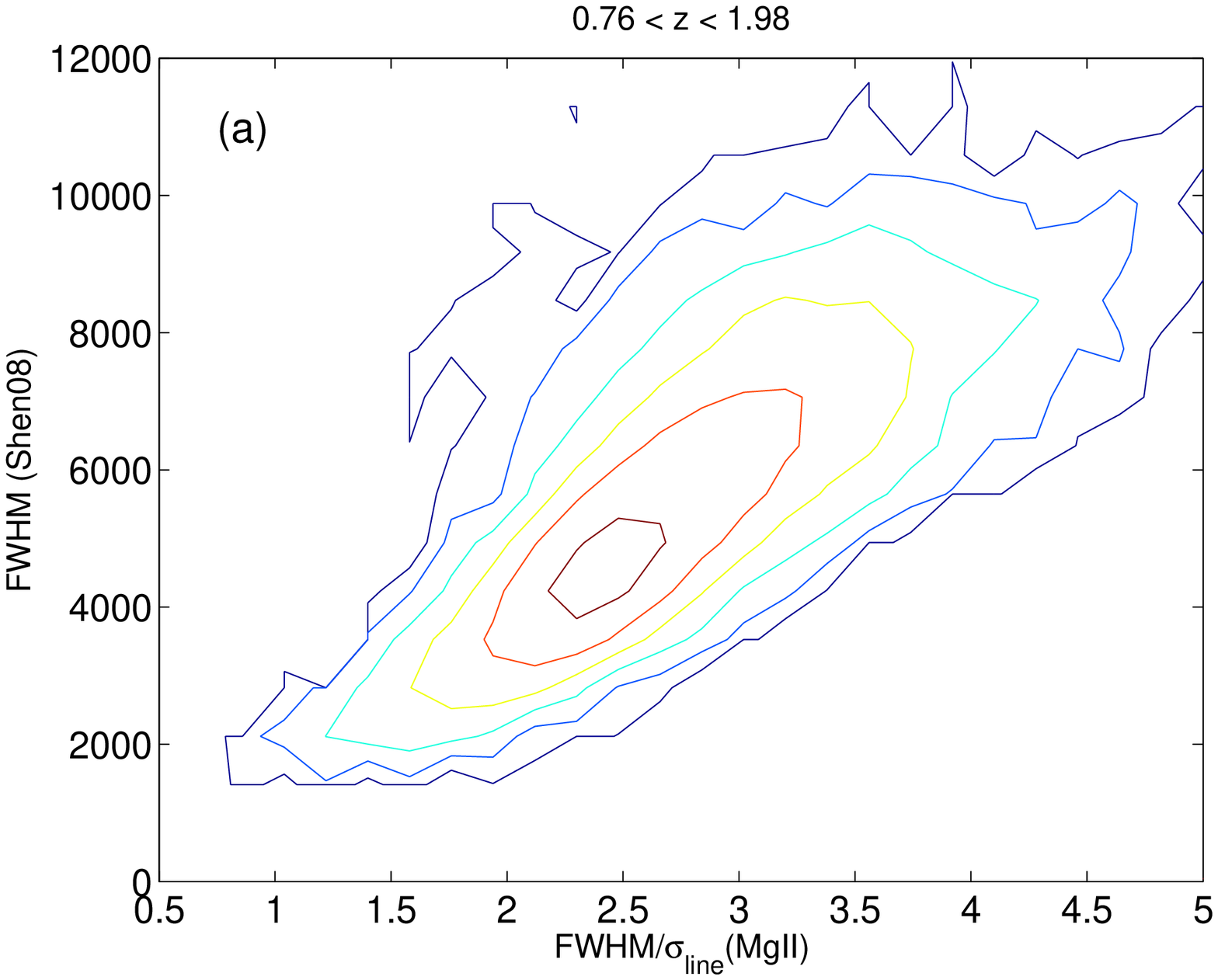}
  \includegraphics[width=3in,height=2.5in]{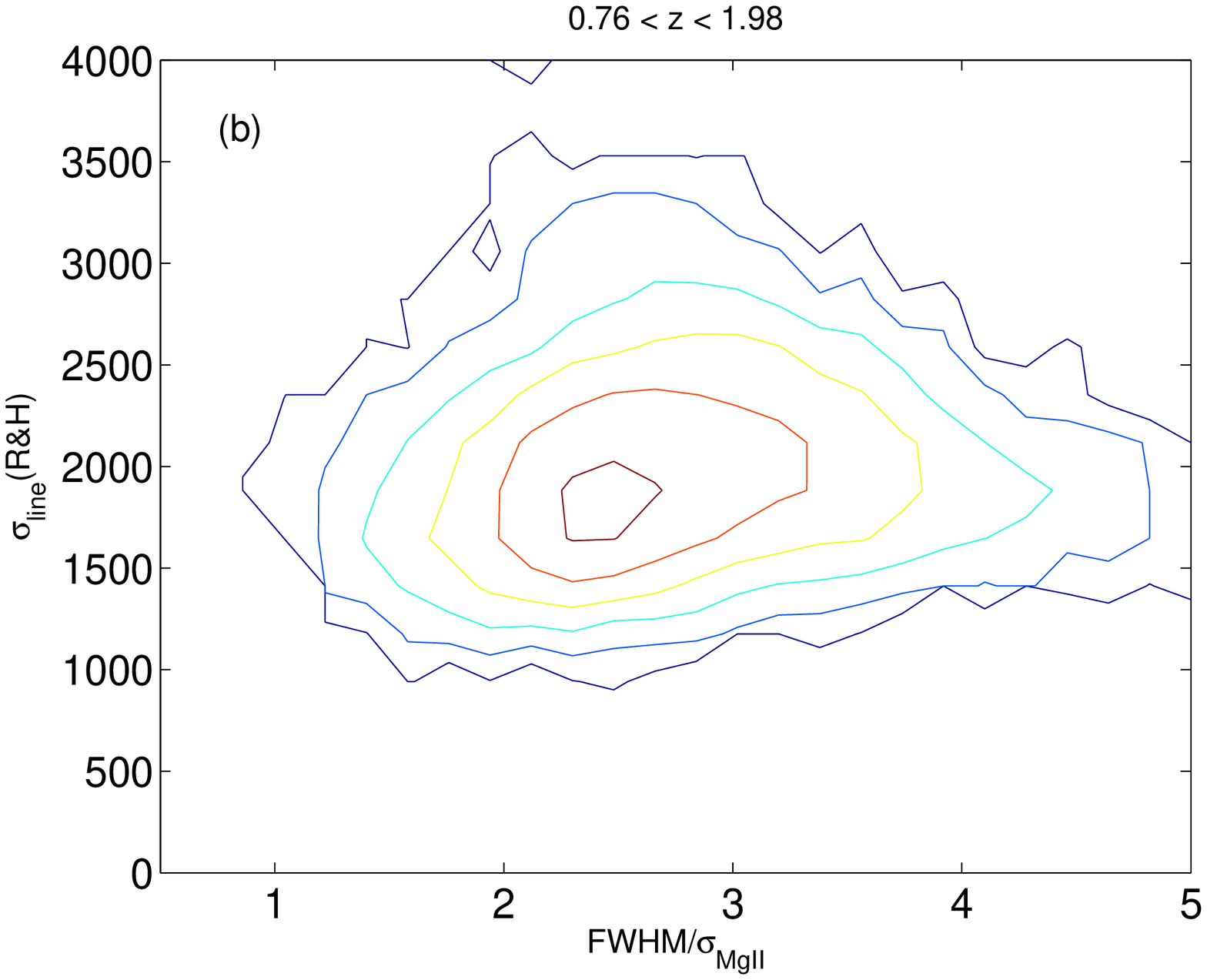}
  \includegraphics[width=3in,height=2.5in]{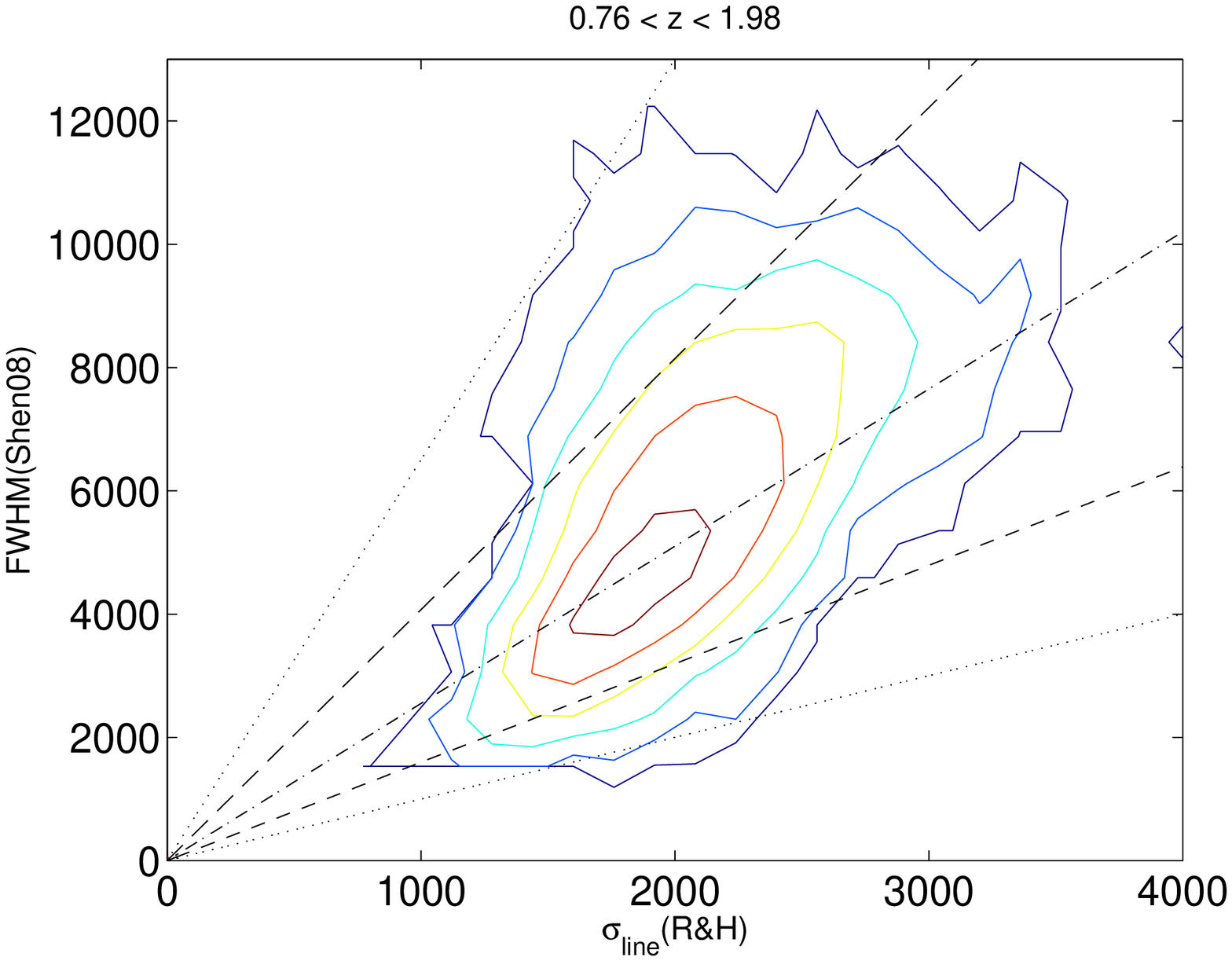}
    \includegraphics[width=3in,height=2.5in]{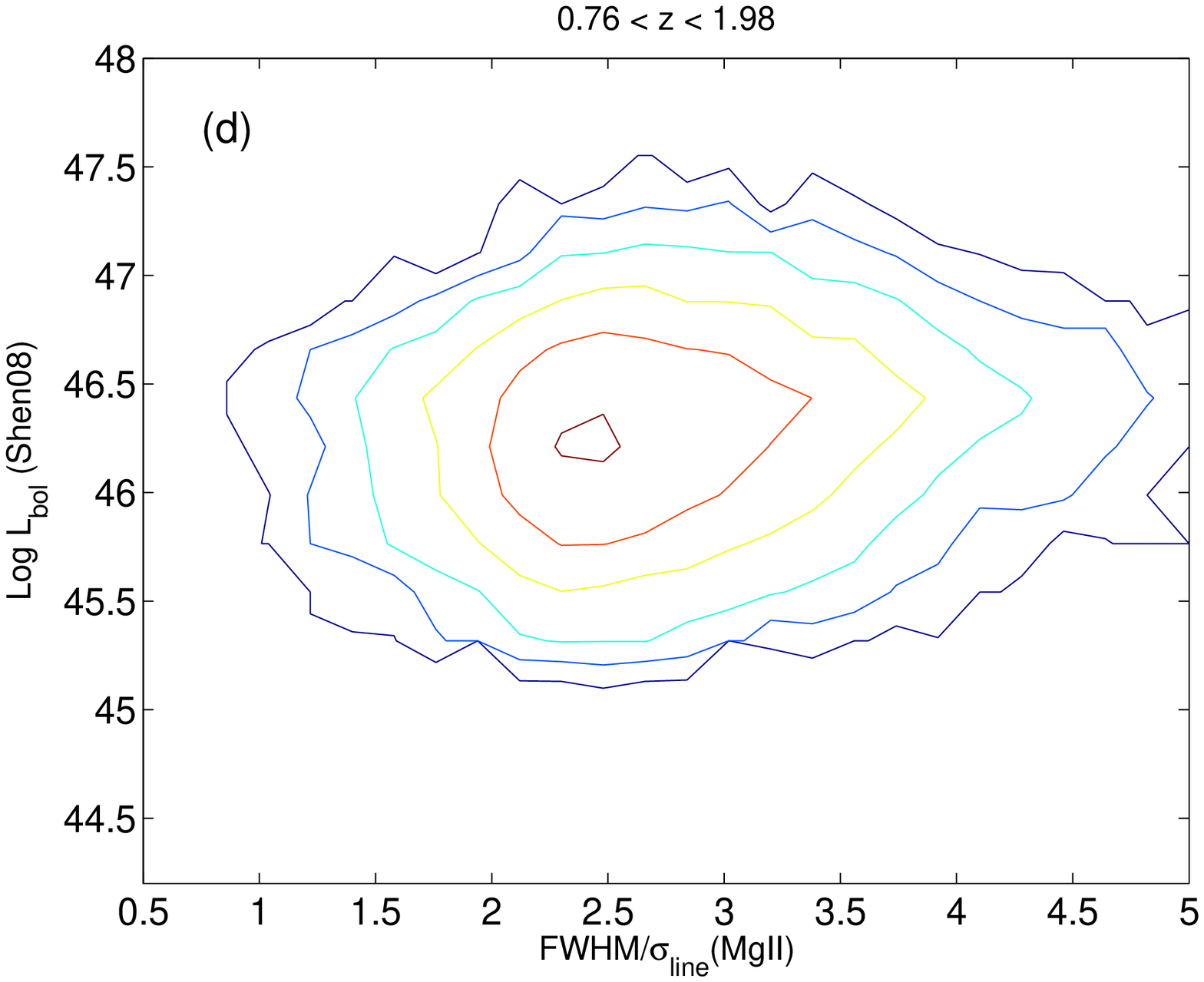}\\
   \caption{Contour plots of the distributions of two \MgII\ line width indicators versus their ratio $\textrm{FWHM}/{\sigma_{line}}$.  The label above each plot gives the redshift range of the sample shown.  a) Using FWHM from Shen et al. (2008) as the line width indicator.  b) Using $\sigma_{line}$ from Rafiee \& Hall (2011) as the line width indicator.  c) Contour plot of the FWHM versus $\sigma_{line}$; the dotted lines show FWHM $=(2.55)^{n/2}\times\sigma_{line}$ for $n=0$ and $n=4$, dashed lines show the same relation for $n=1$ and $n=3$, dot-dashed line for $n=2$.  At any given $\sigma_{line}$, the distribution of FWHM/$\sigma_{line}$ between the five reference lines is roughly the same, while as a function of FWHM the FWHM/$\sigma_{line}$ distribution varies much more substantially. d) Contour plot of the log of the bolometric luminosity from Shen et al. (2008) versus the ratio $\textrm{FWHM}/{\sigma_{line}}$.
}\label{Fig:contour_FWHM2sigmaratio_vs_Shen_RH}
\end{figure*}

In Figures \ref{Fig:contour_FWHM2sigmaratio_vs_Shen_RH}a and \ref{Fig:contour_FWHM2sigmaratio_vs_Shen_RH}b, we show contour plots of the distribution of these two \MgII\ line width indicators as a function of their ratio.
There is a strong correlation of FWHM with the ratio FWHM/$\sigma_{line}$,
but no significant correlation for $\sigma_{line}$.
Figure \ref{Fig:contour_FWHM2sigmaratio_vs_Shen_RH}c plots FWHM versus $\sigma_{line}$. This contour plot is another way of demonstrating that the distribution of FWHM/$\sigma_{line}$ (e.g., between the five reference lines plotted) as a function of $\sigma_{line}$ shows only moderate variation from an average distribution, while as a function of FWHM the distribution of FWHM/$\sigma_{line}$ shows great variation.
Recall that Peterson et al. (2004) found that while the virial velocity in the
BH mass relation is well characterized by $\sigma_{line}$, it is less well
characterized by FWHM (\S 1).
Combining that result and ours, we can say that $\sigma_{line}$ is a good
representation of the virial velocity regardless of the line shape
(as crudely measured by FHWM/$\sigma_{line}$), whereas FWHM is not.
This change in the typical ratio between FWHM and $\sigma_{line}$ with FWHM
again calls into question the use of a BH mass relation calibrated to the FWHM.

\begin{figure*}
  \includegraphics[width=3in,height=2.5in]{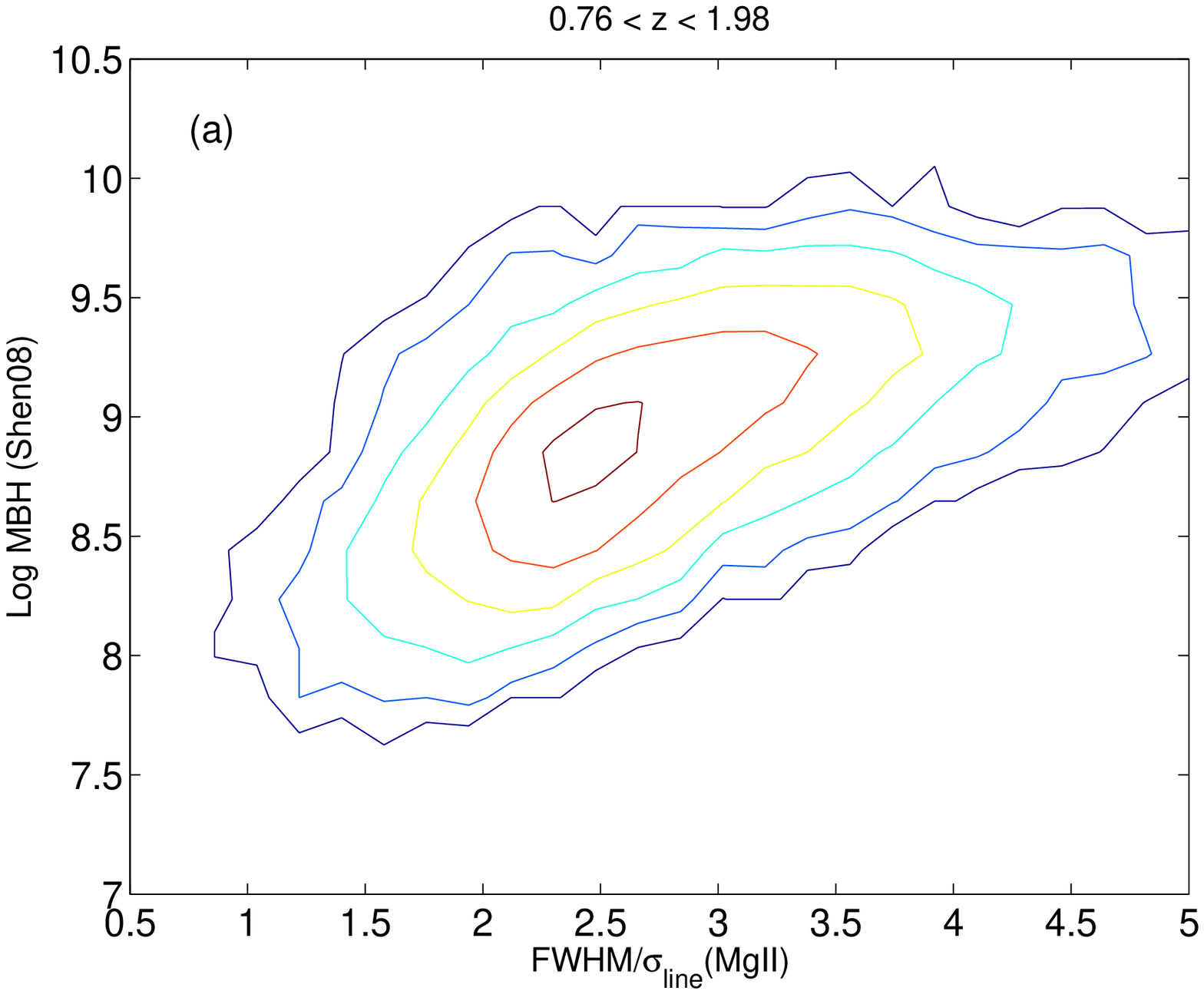}
  \includegraphics[width=3in,height=2.5in]{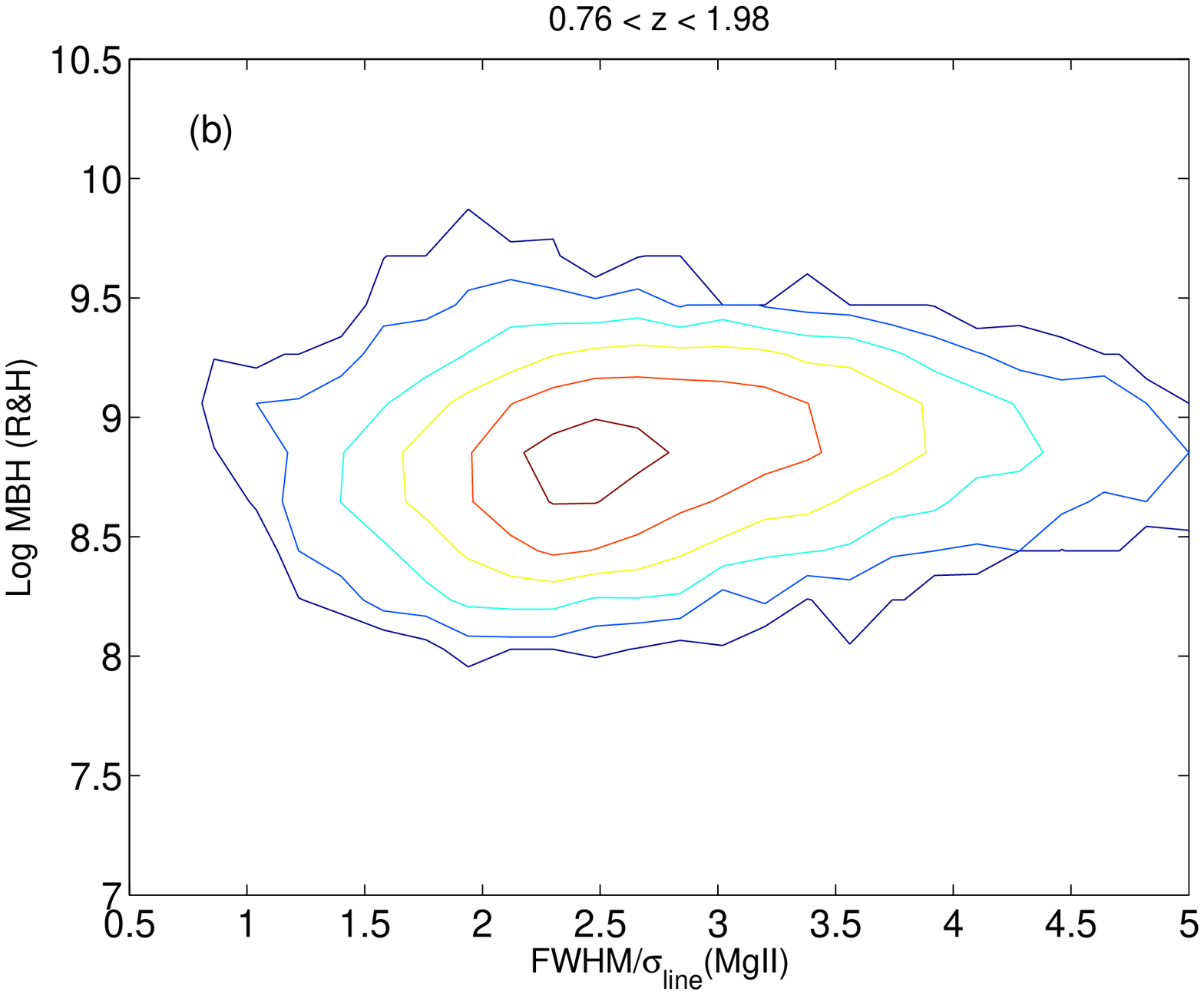}
  \includegraphics[width=3in,height=2.5in]{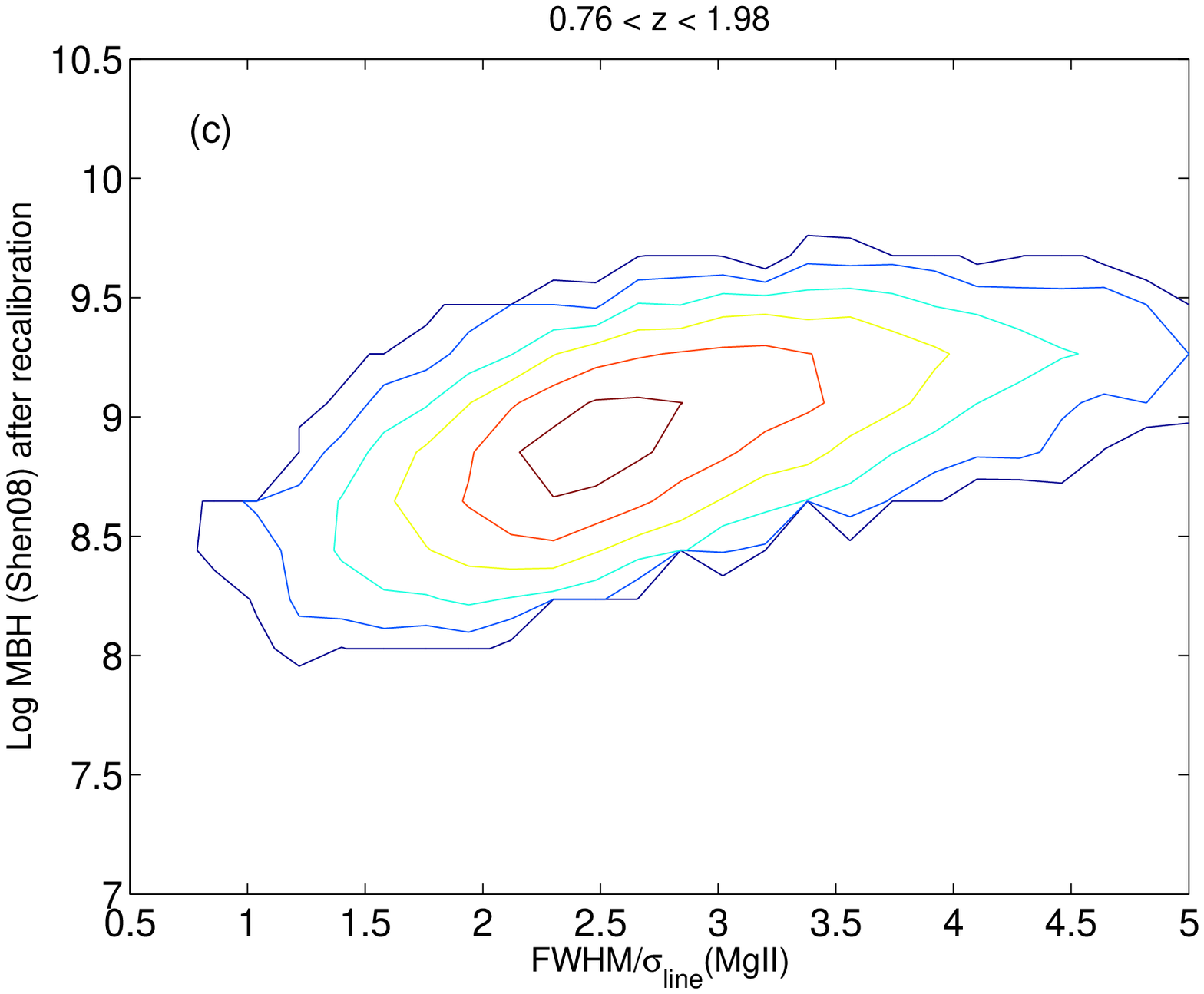}\\
   \caption{Contour plot of the distribution of log \MgII\ black hole mass versus $\textrm{FWHM}/{\sigma_{line}}$ ratio. a) Using Shen et al. (2008) \MgII\ BH mass estimates (before recalibration). b) Using Rafiee \& Hall (2011) \MgII\ mass estimates. c) Using the FWHM from Shen et al. (2008) to estimate the BH masses via Equation \ref{equ:new_mass_FWHM}.
   In all plots, the FWHM/$\sigma_{line}$ ratio uses FWHM from Shen et al. (2008) and $\sigma_{line}$ from Rafiee \& Hall (2011).}\label{Fig:contour_FWHM2sigmaratio_vs_Shen_RH_BHmass}
\end{figure*}

In Figure \ref{Fig:contour_FWHM2sigmaratio_vs_Shen_RH_BHmass}, we show contour plots of quasar BH masses versus ${\rm FWHM}/\sigma_{line}$ for three scenarios. Panel (a) shows the Shen et al. (2008) results which assume a Gaussian profile for \MgII, and panel (b) the Rafiee \& Hall (2011) results which directly use the line dispersion.
Figure \ref{Fig:contour_FWHM2sigmaratio_vs_Shen_RH_BHmass}a shows that the dependence of FWHM on the FWHM/$\sigma_{line}$
ratio means that a broader range of BH masses are found using FWHM-based
estimation as compared to $\sigma_{line}$-based estimation (Figure \ref{Fig:contour_FWHM2sigmaratio_vs_Shen_RH_BHmass}b).
Figure \ref{Fig:contour_FWHM2sigmaratio_vs_Shen_RH_BHmass}c shows that the recalibration of FWHM-based masses discussed in
the next section helps to reduce the discrepancy between the range of BH
masses estimated using FWHMs and that estimated using $\sigma_{line}$ values.

\section{Recalibrating the Black Hole Mass Scaling Relation}\label{sec:BH_Mass_rescaling}

\subsection{Statistical Methods}\label{sec:stats}

We have used numerous different methods to estimate the best-fit parameters of two- and three-parameter BH mass scaling relationships.
However, we base our conclusions on the MLINMIX\_ERR method (Kelly et al. 2007), also used by Wang et al. (2009), which we report in Table \ref{tab:Bentz_Wang_MJ_3logfit}.
The MLINMIX\_ERR method is preferred over other methods because it is the only fitting method which takes intrinsic scatter and uncertainties in both parameters into account. It should be the default fitting method in such cases, though its results may not differ significantly from other methods for two-parameter fits. We also use another Metropolis-Hastings Markov chain Monte Carlo simulation (MCMC; Haario et al. 2006) which gives very similar results to LINMIX\_ERR and MLINMIX\_ERR; they both use the Bayesian approach to the regression problem but use different sampling algorithms. The MCMC and MLINMIX\_ERR (and LINMIX\_ERR) simulations yield confidence levels as well as potential outliers and the distribution of the acceptable parameters and their associated errors within the parameter space.

Other methods used in previous studies are less sophisticated. For example, the BCES method (Akritas \& Bershady 1996, e.g., used by McLure \& Dunlop 2004, Kaspi et al. 2005, Vestergaard et al. 2006) considers bivariate correlated errors and a possible intrinsic scatter but it does not report that scatter. A quantified value of the intrinsic scatter is available from the FITexy-T02 method (Tremaine et al. 2002, e.g., used by Kaspi et al. 2005, Vestergaard et al. 2006). However, FITexy-T02 does not account for the bivariate correlated errors in the fitting process. Nonetheless, for convenience
and to illustrate the large dispersion which results from applying non-optimal fitting algorithms,
we report the results from all these methods in Table \ref{tab:Re_Fitting_Results_Bentz}.

\subsection{The Wang et al. Three-Parameter Relationship}\label{sec:wang3param}

Wang et al. (2009) have studied the mass scaling relations for the \Hbeta\ and \MgII\ emission lines in 29 out of 35 low redshift AGNs for which Peterson et al. (2004) have reported mass estimates made through reverberation mapping (RM) (see Table \ref{tab:Bentz_Wang_MJ_3logfit}).
Because the single-epoch FWHM of a given line is unlikely to be an exact tracer of
its virial line width,
in determining new mass-scaling relations they have assumed
a more flexible relation between the virial line width and the FWHM
in the form of $v^2_{virial}\propto \textrm{FWHM}^{\gamma}$
where $\gamma$ is not fixed at 2, as in conventional virial relations,
but rather is a free parameter.

In their mass scaling, Wang et al. (2009) adopted the RM BH masses given in the Peterson et al. (2004) for 24 objects but used the RM BH masses from more recent reverberation campaigns for the other 5 objects: NGC 4593 from Denney et al. (2006), NGC 4151 from Metzroth et al. (2006), PG 2130+099 from Grier et al. (2008), and NGC 4051 from Denney et al. (2009).  They (and we) take the $\sigma_{line}$(\Hbeta,rms) values of all 29 RM objects from Peterson et al. (2004).

For their full 29-quasar RM sample, Wang et al. (2009) found a different power-law relation between FWHM(\MgII) and $\sigma_{line}$(\Hbeta,rms):
\begin{eqnarray}\label{Equ:FWHM_MgII_Hbeta}
 \log\left[\frac{\sigma_{line}(\textrm{\Hbeta},\textrm{rms})}{1000~\textrm{km s}^{-1}}\right]=(0.85\pm0.21)\log\left[\frac{\textrm{FWHM}(\textrm{\MgII})}{1000~\textrm{km s}^{-1}}\right]
   \nonumber\\
 -(0.21\pm0.12).
\end{eqnarray}
Based on this result, Wang et al. (2009) concluded that the line-emitting locations of \MgII\ are different from those of \Hbeta\ in the broad line region (BLR).  That conclusion may be premature, as the difference from unity slope is $< 1\sigma$ statistical significance and the FWHM(\MgII) values are measured from single epoch spectra and not from the same rms spectra as the $\sigma_{line}$(\Hbeta) values.  Nonetheless, they suggest fitting a three parameter relation between \mbh, FWHM and $\lambda L_{3000}$:
\begin{eqnarray}\label{Equ:Wang_3param_scaling}
\log\left[\frac{\textrm{\mbh}(RM)}{10^6\textrm{\msol}}\right] = \alpha + \beta\times \log\left[\frac{\lambda L_{\lambda}}{10^{44}~\textrm{erg s}^{-1}}\right]
   \nonumber\\
 + \gamma\times\log\left[\frac{\textrm{FWHM}(\textrm{\MgII})}{1000~\textrm{km s}^{-1}}\right]
\end{eqnarray}
where $\gamma=1.7\pm0.42$ provides consistency between
\MgII-based single-epoch BH masses and \Hbeta-based RM BH masses.

\subsection{An Up-to-date, Recalibrated Three-Parameter Relationship}\label{sec:our3param}

Motivated by Wang et al. (2009), we have revisited the mass scaling relation used by Shen et al. (2008), which was derived by McLure \& Dunlop (2004) based on 20 RM objects studied in McLure \& Jarvis (2002). McLure \& Dunlop (2004) reported a one-to-one relation between FWHM-based \mbh\ (\Hbeta) and \mbh\ (\MgII), $\log\textrm{\mbh}(\textrm{\Hbeta})=1.00(\pm0.08)\log\textrm{\mbh}(\textrm{\MgII})+0.06(\pm0.46)$ and concluded that \Hbeta\ can be replaced by \MgII\ for purposes of mass estimation.

We have updated the BH mass list used by Wang et al. (2009) with the most recent RM BH mass estimates of NGC 5548 (Bentz et al. 2009) and NGC 3227 (Denney et al. 2010).
Using those updated BH masses,
we have recalibrated the Wang et al. (2009) mass relation.
Comparing results before and after updating the RM subsample shows no statistically significant change in the fitting results, except in those of FITexy and
some improvement in the value of the correlation coefficient in the
three-parameter fitting results.
This improvement means that the regression line fits the updated sample better than that of Wang et al. (2009) or McLure \& Jarvis (2002).

Here, we report the three parameter mass scaling relations using the most updated RM masses and the MLINMIX\_ERR regression package.
\begin{eqnarray}\label{equ:new_mass_FWHM}
\log\left[\frac{\textrm{\mbh}(RM)}{10^6\textrm{\msol}}\right] = (1.25\pm0.22)
    + (0.51\pm0.08)
    \nonumber\\
    \times \log\left[\frac{\lambda L_{\lambda}}{10^{44}~\textrm{erg s}^{-1}}\right]
    \nonumber\\
    + (1.27\pm0.40)\times\log\left(\frac{\textrm{FWHM}(\textrm{\MgII})}{1000~\textrm{km s}^{-1}}\right)
   \nonumber\\
   \pm\sigma_{\log[\textrm{\mbh}/10^6\textrm{\msol}]}(\textrm{statistical})
   \nonumber\\
   \pm(0.15\pm0.5)~\textrm{dex} (\textrm{intrinsic~scatter}).
\end{eqnarray}
where $\sigma_{\log[\textrm{\mbh}/10^6\textrm{\msol}]}$(statistical) is the statistical error of the black hole mass from:
\begin{eqnarray}\label{equ:new_FWHM_MBH}
    \sigma_{\log[\textrm{\mbh}/10^6\textrm{\msol}]}=  [ 0.048+ 0.0012\, \left( \ln  \left( {\ell} \right)
 \right) ^{2}+ 0.048\,{\frac {{\sigma_{{{\ell}}}}^{2}}{{{\ell}}^
{2}}}
    \nonumber\\
+ 0.030\, \left( \ln  \left( {\omega} \right)  \right) ^{2}+
 0.32\,{\frac {{\sigma_{{{\omega}}}}^{2}}{{{\omega}}^{2}}}] ^{ 0.5}
\end{eqnarray}
where $\ell=[\lambda L_{3000}/10^{44}]$ is in units of \ergps, $\omega=[\textrm{FWHM}/1000]$ is in units of \kmps, and $\sigma_{\omega}$ and $\sigma_{\ell}$ are estimated errors from our fitting process. Like Wang et al. (2009), we find that the relationship of \mbh\ to \MgII\ FWHM follows a power law with a smaller value than 2 but that the deviation is not statistically significant ($< 2\sigma$).


\subsection{Recalibrated Mass-Luminosity Plane}\label{sec:recalibrated_mass_luminosity_plane}

\begin{figure*}
  \includegraphics[width=5in,height=4in]{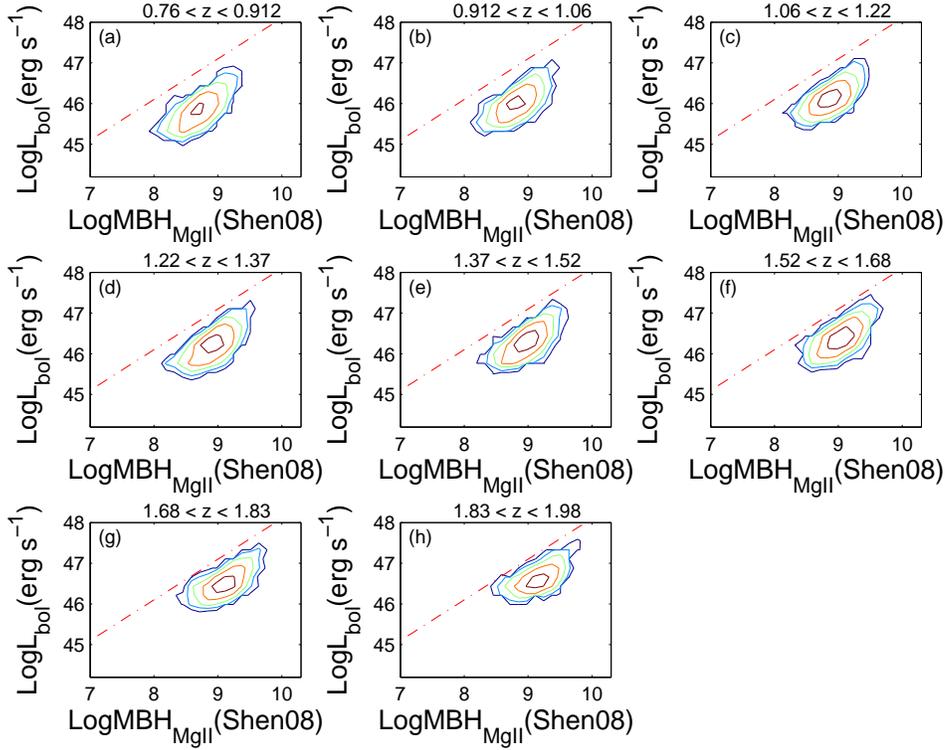}\\
 \caption{The quasar \MgII\ mass-luminosity plane for 8 redshift bins when we apply a new mass-scaling relation to the original FWHM and luminosity of Shen et al. (2008).
The slope of the upper envelope of the quasar distribution is consistent with unity in all redshift bins.}\label{Fig:8quasar_mass_luminosity_plane_rescale_newshen}
\end{figure*}

We have shown in Figure \ref{Fig:8in1_zbin_Lbol_MBH_RH_calibration} that
when the Rafiee \& Hall (2011) sample shows a SEB in the mass-luminosity plane, its slope parallels that of the Eddington limit, unlike what was found by SE10a.  The same is true with most of the new mass scaling relations.
The mass-luminosity plane using the Shen et al. (2008) measurements with a re-scaled mass-relation assuming $v^2_{virial} \propto \textrm{FWHM}^{\gamma}$ with $\gamma < 2$ shows no sign of a SEB with non-unity slope (see Figure \ref{Fig:8quasar_mass_luminosity_plane_rescale_newshen}).
However, the slope of the SEB may be less than unity if $\gamma \geq 2$; as a change from $\gamma=2$ to $\gamma<2$ has increased the slope of the SEB, a change from $\gamma=2$ to $\gamma>2$ should decrease its slope.

\begin{figure*}
  \includegraphics[width=3in,height=2.5in]{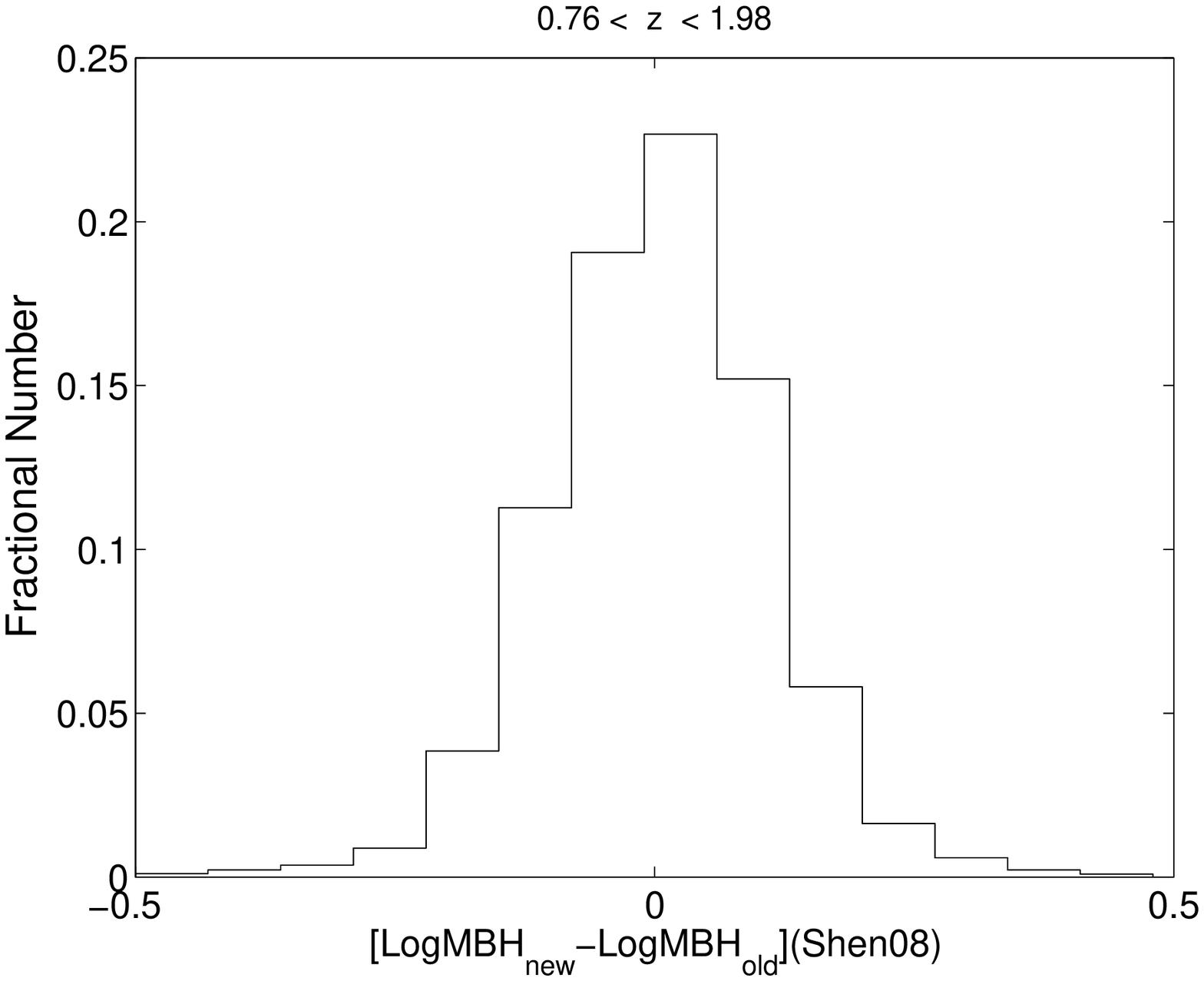}
  \includegraphics[width=3in,height=2.5in]{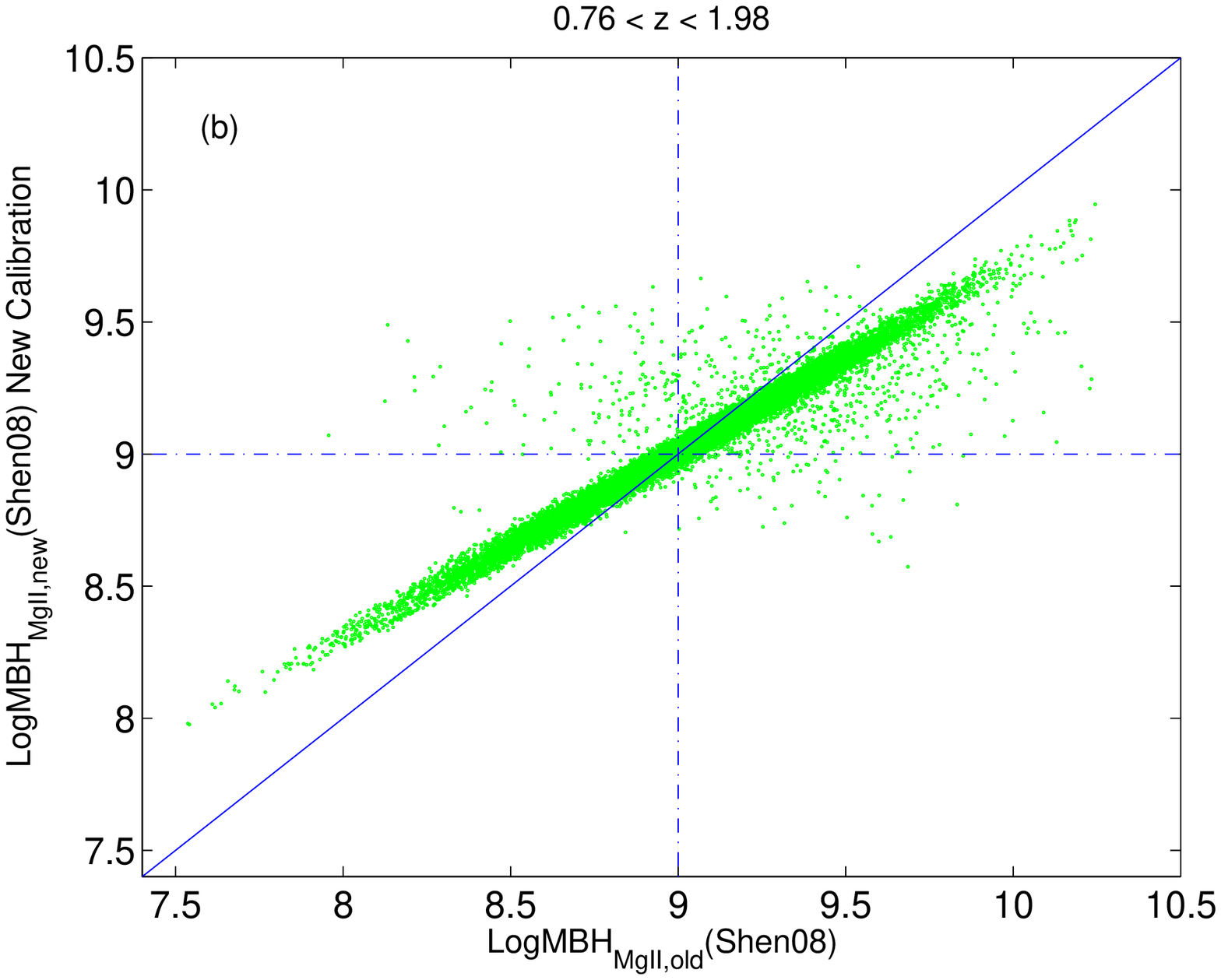}\\
  \includegraphics[width=3in,height=2.5in]{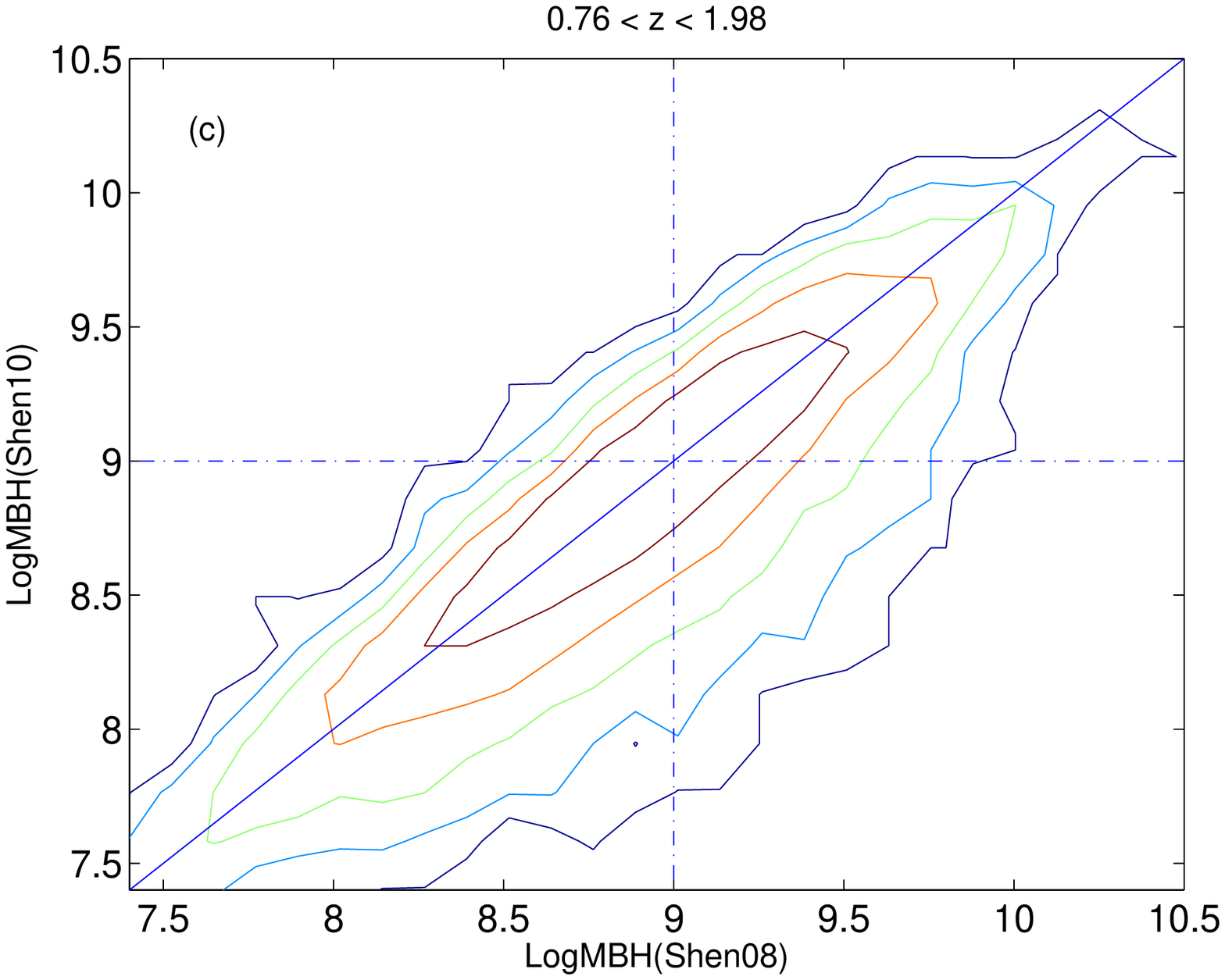}\\
   \caption{a) The log difference between the Shen et al. (2008) BH masses using the old and new calibrations. There is no systematic shift in the mass distribution. b) Scatter plot of the same BH masses. The mass distribution has rotated around the value $\log \textrm{\mbh} \sim 9$ under the new calibration, causing the SEB to disappear. c) Contour plot of BH mass estimates from Shen et al. (2010) versus Shen et al. (2008) for the \MgII\ redshift range.
}\label{Fig:MBH_Shen_histogram_Newnonlin_vs_old_calibration}
\end{figure*}

Our results show that a recalibration of the mass scaling relation can produce different shifts in the high and low mass tails of the distribution, rather than a uniform overall shift to lower or higher masses.
Compared to masses recalibrated with the latest mass-scaling relation, the published Shen et al. (2008) masses overestimate the highest BH masses and underestimate the lowest BH masses. The overall distribution of the log difference of mass is nearly symmetric around zero, as shown in Figure \ref{Fig:MBH_Shen_histogram_Newnonlin_vs_old_calibration}a. However, the new mass estimates versus the Shen et al. (2008) masses depicted in Figure \ref{Fig:MBH_Shen_histogram_Newnonlin_vs_old_calibration}b show a rotation in the mass distribution around $\log \textrm{\mbh}\sim 9$ instead of a systematic shift in the new mass estimates.
This rotation around $\log \textrm{\mbh}\sim 9$ (Figure \ref{Fig:MBH_Shen_histogram_Newnonlin_vs_old_calibration}b) may explain the presence of a SEB of non-unity slope in the original Shen et al. (2008) data, especially at the highest masses.

Another potential cause of differences between BH mass datasets is the value of $\beta$ used for the luminosity dependence in the mass calibration equation.  The SE10a mass scaling relations in \Hbeta, \MgII\ and \CIV\ used $\beta=0.61$, $0.62$ and $0.53$, respectively, whereas for \MgII\ we use $\beta=0.5$ in Rafiee \& Hall (2011) and $\beta=0.51$  for the recalibrated three-parameter relationship above.  Even assuming a perfect correlation between FWHM and $\sigma_{line}$, for two quasars of luminosity differing by a factor of 10 for which our method measures the same mass, SE10a would obtain a mass higher by 0.11 or 0.12 in the log for the higher luminosity quasar.  That systematic difference will decrease the slope of the upper envelope of the quasar distribution in the mass-luminosity plane, although not by enough to explain the full deviation from unity of the SEB slope in SE10a.

\subsection{Other Surveys}\label{sec:other}

Other surveys which use the FWHM as a line width indicator may yield biased
BH masses if $\gamma=2$ is assumed.

Kollmeier et al. (2006) use the FWHMs of \Hbeta, \MgII\ and \CIV\ measured
directly from smoothed spectra from the AGES survey for their BH mass
measurements at $0.3<z<4$, with
$\gamma=2$ and $\beta=0.88$ for \MgII\ at $0.4<z<2$.
They do not report a SEB of non-unity slope, but may see a weak one
in their $1<z<2$ redshift bin (their Figure 10, lowest three panels).

Netzer \& Trakhtenbrot (2007) use the FWHM of the sum of two broad Gaussians
fit (along with one narrow Gaussian) to \Hbeta\ in SDSS spectra for their
BH mass measurements at $z<0.75$.
They use a mass scaling relation with $\gamma=2$ but $\beta=0.65$,
and appear to find a SEB of non-unity slope (their Figure 1, bottom panel,
and Figure 3, red and blue histograms).

The high value of $\beta$ in the above calibrations may partially compensate for any biases due to the use of FWHM or its calibration.  The monochromatic luminosity is little affected by the shape of the line (see Figure \ref{Fig:contour_FWHM2sigmaratio_vs_Shen_RH}d), while the FWHM clearly depends on the shape (see e.g., Figure \ref{Fig:contour_FWHM2sigmaratio_vs_Shen_RH}a).  A heavy weight given to the luminosity in the scaling relationship reduces the effect of the sensitivity of the FWHM to the line shape.  This may reduces the presence of the SEB in, e.g., Kollmeier et al. (2006).

\subsection{Results using the SDSS Seventh Data Release}\label{sec:results_using_data_release_seven}
\begin{figure*}
  \includegraphics[width=5in,height=4in]{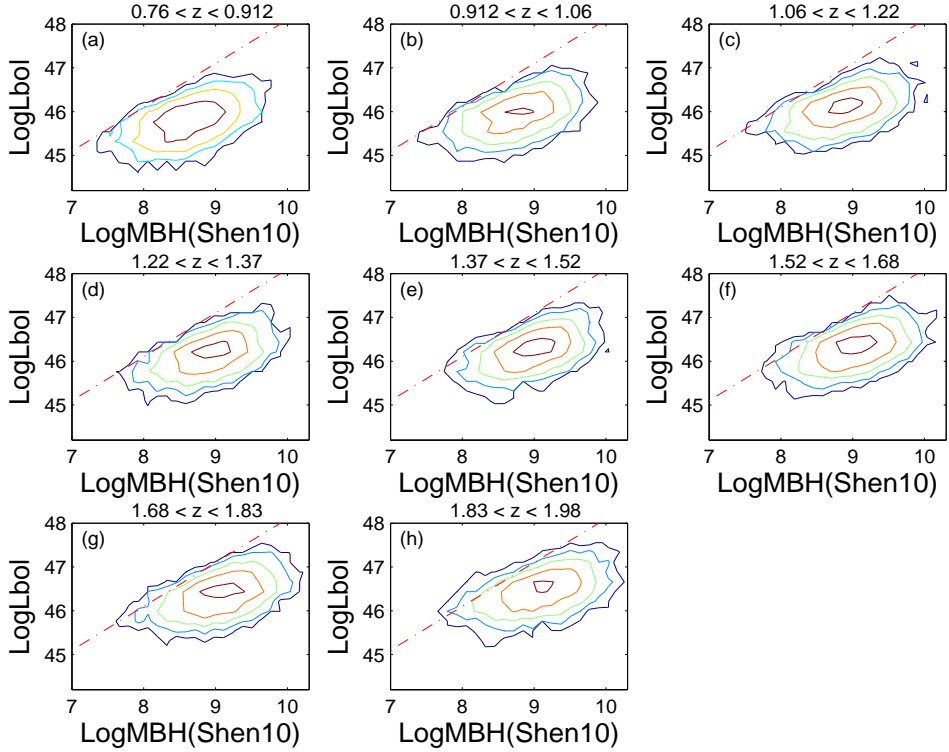}\\
 \caption{ The quasar \MgII\ mass-luminosity plane for 8 redshift bins using the fiducial ``S10'' BH masses from Shen et al. (2010). A sub-Eddington boundary of non-unity slope for objects with the highest \mbh\ in a given redshift bin can be seen partially in panels (a) to (e), as found by SE10a.}\label{Fig:8quasar_mass_luminosity_plane_Shen10}
\end{figure*}

\begin{figure*}
  \includegraphics[width=5in,height=4in]{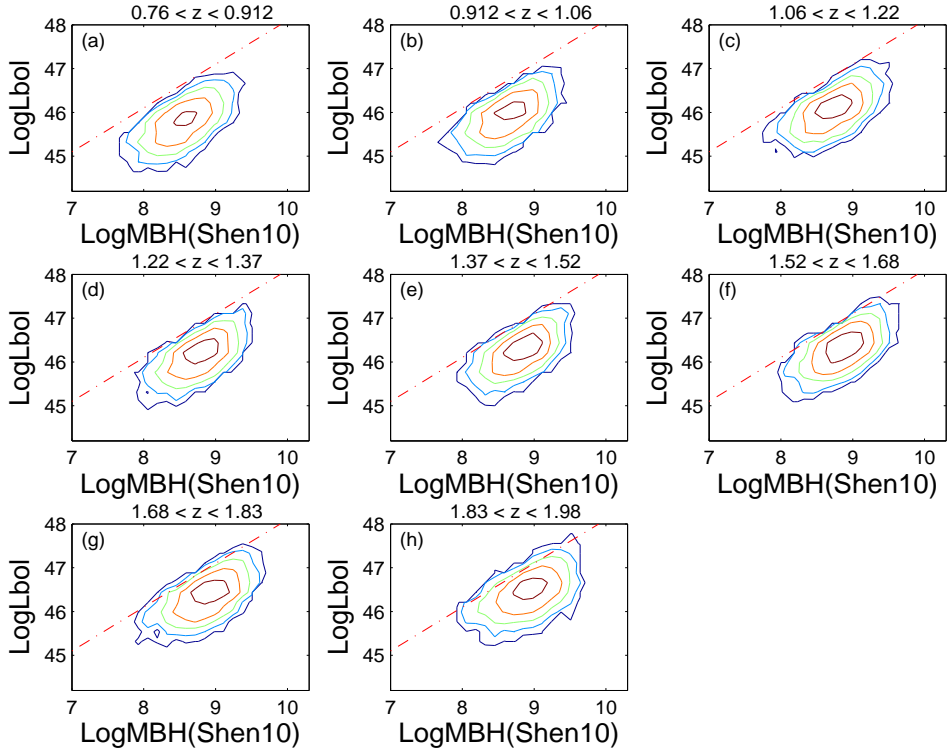}\\
\caption{ The quasar \MgII\ mass-luminosity plane for 8 redshift bins using Equation \ref{equ:new_mass_FWHM} to re-estimate the ``S10'' BH masses from Shen et al. (2010). No sub-Eddington boundary of any form, with non-unity or unity slope, can be seen here. }\label{Fig:8quasar_mass_luminosity_plane_Shen10_new}
\end{figure*}

Shen et al. (2010) have reported estimates for BH masses of 105783 quasars in the SDSS Seventh Data Release (DR7).
Their fiducial (``S10'') \MgII-based BH mass estimates use the FWHM of the broad component of \MgII\ as fitted by up to 3 Gaussians in conjunction with a narrow Gaussian of width $\leq$1200~\kmps.
These mass estimates are calibrated by adopting the McLure \& Dunlop (2004) slope $\beta=0.62$ and setting the normalization $\alpha$ to match the H$\beta$-based masses of Vestergaard \& Peterson (2006) on average.
Figure \ref{Fig:MBH_Shen_histogram_Newnonlin_vs_old_calibration}c shows that the BH mass estimates from Shen et al. (2010) have not significantly changed with respect to the mass catalogue reported by Shen et al. (2008).
In Figure \ref{Fig:8quasar_mass_luminosity_plane_Shen10}, we show the results of using this larger sample with updated \MgII-based mass estimates.  Furthermore, following Wang et al. (2009), we can use the recalibrated mass-scaling relation reported in Equation \ref{equ:new_mass_FWHM} to recalculate the BH masses of objects within the redshift range $0.76 < z < 1.98$.  The results, shown in Figure \ref{Fig:8quasar_mass_luminosity_plane_Shen10_new}, show no SEB of any kind in the mass-luminosity plane.
Apparently, the presence of more low luminosity quasars in DR7 along with the correction applied to the mass-scaling relation have removed the signature of the SEB from the mass-luminosity plane.

\section{Conclusion}\label{sec:Conclusion_SEB}

SE10a have argued that the presence of a sub-Eddington boundary (SEB) with non-unity slope in the mass-luminosity plane constructed from the Shen et al. (2008) FWHM-based BH mass data set is not due to measurement error and is statistically significant. They have shown that the SEB they found is not an artifact due to small number statistics.
We have investigated the existence of such a SEB using our own data set (Rafiee \& Hall 2011) which uses the line dispersion of \MgII\ to estimate BH masses. Furthermore, motivated by Wang et al. (2009), we have considered the effects of recalibrating the mass scaling relationship using the general three-parameter relation proposed in Equation \ref{Equ:Wang_3param_scaling} and taking into account the latest mass updates from reverberation campaigns. Finally, we have investigated the SEB using the larger SDSS DR7 BH mass catalogue of Shen et al. (2010). Overall, we conclude that:
\begin{itemize}
\item
We have found no sign of a SEB with non-unity slope in the Rafiee \& Hall (2011) data set, before or after recalibration, nor in the recalibrated Shen et al. (2008) or Shen et al. (2010) data sets.

\item
The non-unity slope of the SEB in the original data of Shen et al. (2008) arose from the mass-scaling relation used and is likely \emph{not} a real feature in the quasar mass-luminosity plane.
However, the implications of a sub-Eddington boundary of unity slope remains
a worthwhile topic for theoretical investigation.

\item
The presence or absence of a SEB of non-unity slope is rooted in the choice of line width indicator and in the calibration of the mass scaling relation used. Further studies of single epoch spectra of RM sample AGN may give clues about the roots of the problem with different line width indicators and calibrations. Until then, we suggest using the line dispersion, which is a more robust parameter to represent the line width, to estimate the BH masses of quasars when using a scaling relation of the form $v_{virial}\propto\sigma_{line}$. The FWHM can be used; however, the scaling relation may have a different form $v_{virial}^2\propto{\rm FWHM}^\gamma$, with $\gamma<2$.

\end{itemize}

We thank the referee, B. M. Peterson, for his valuable comments and discussions. PBH and AR are supported in part by NSERC. Funding for the SDSS and SDSS-II was provided by the Alfred P. Sloan Foundation, the Participating Institutions, the National Science Foundation, the U.S. Department of Energy, the National Aeronautics and Space Administration, the Japanese Monbukagakusho, the Max Planck Society, and the Higher Education Funding Council for England. The SDSS was managed by the Astrophysical Research Consortium for the Participating Institutions.

\begin{table*}
 \centering
  \begin{minipage}{155mm}
  \caption{Three-parameter fitting results using MLINMIX\_ERR (reporting median posterior distribution). $\log(\textrm{\mbh}/10^6\textrm{\msol})$ = $\alpha$ + $\beta\times \log(\lambda L_{\lambda}/10^{44}$\textrm{ergs$^{-1}$}) + $\gamma\times\log(FWHM/1000$\textrm{kms$^{-1}$}).\label{tab:Bentz_Wang_MJ_3logfit}}
  \begin{tabular}{@{}lccccccc@{}}
\hline
\textbf{Data resource}& Number of objects & line width &$\alpha$& $\beta$ & $\gamma $
&  & \textbf{Intrinsic Scatter} \\
\hline
Wang et al. 2009 & 29 & FWHM &	1.26	$\pm$	0.23	&	0.46	$\pm$	0.08	&	1.28	$\pm$	0.42	&		 &	 0.16	$\pm$	0.06	 \\
This study  & 29 & FWHM &	1.25	$\pm$	0.22	&	0.51	$\pm$	0.08	&	1.27	 $\pm$	 0.40	&	&	0.15	 $\pm$	0.06	\\
(with latest updates & & & & & & &\\
on RM masses) & & & & & & &\\
\hline
\end{tabular}
\end{minipage}
\end{table*}
\begin{table*}
 \begin{minipage}{100mm}
  \caption{Re-scaling results using most updated black hole masses. $\log(\textrm{\mbh}/10^6\textrm{\msol})-0.5\log(\lambda L_{\lambda}/10^{44}$\textrm{ergs$^{-1}$}) = $\alpha$ + $\gamma\times\log(FWHM/1000$\textrm{kms$^{-1}$}).\label{tab:Re_Fitting_Results_Bentz}}
  \begin{tabular}{@{}lcccccc@{}}
\hline
\textbf{Fitting Method}& Intercept($\alpha$)& Slope($\gamma$) & \textbf{Intrinsic Scatter} & \\
\hline
OLS(Y$|$X)	&	1.244	$\pm$	0.25	&	1.27	$\pm$	0.47	&				\\
OLS(X$|$Y)	&	-0.136	$\pm$	0.38	&	3.92	$\pm$	0.66	&				\\
OLS Bisector	&	0.85	$\pm$	0.24	&	2.03	$\pm$	0.42	&				\\
OLS Bisector bootstrap	&	0.833	$\pm$	0.26	&	2.05	$\pm$	0.44	&				\\
OLS Bisector Jacknife	&	0.849	$\pm$	0.29	&	2.03	$\pm$	0.50	&				\\
\hline													
FITexy 	&	0.728	$\pm$	0.09	&	1.99	$\pm$	0.17	&	0			\\
FITexy-T02	&	0.85	$\pm$	0.23	&	1.95	$\pm$	0.43	&	0.29			\\
\hline													
BCES(Y$|$X)	&	1.21	$\pm$	0.27	&	1.34	$\pm$	0.51	&				\\
BCES(Y$|$X) bootstrap	&	1.16	$\pm$	0.33	&	1.41	$\pm$	0.60	&				\\
BCES(X$|$Y)	&	0.0343	$\pm$	0.34	&	3.95	$\pm$	0.58	&				\\
BCES(X$|$Y) bootstrap	&	0.203	$\pm$	202.00	&	3.15	$\pm$	475.00	&				\\
BCES Bisector	&	0.846	$\pm$	0.25	&	2.03	$\pm$	0.44	&				\\
BCES Bisector bootstrap	&	0.822	$\pm$	0.28	&	2.07	$\pm$	0.49	&				\\
BCES Orthogonal	&	0.259	$\pm$	0.31	&	3.16	$\pm$	0.51	&				\\
BCES Orthogonal bootstrap	&	0.346	$\pm$	100.00	&	2.93	$\pm$	235.00	& 			\\
\hline													
MCMC (median posterior distribution)	&	1.24	$\pm$	0.20	&	1.28	$\pm$	0.36	&	0.37	$\pm$	0.05	\\
MCMC (mean posterior distribution)	&	1.24	$\pm$	0.20	&	1.27	$\pm$	0.36	&	0.37	$\pm$	0.05	\\
\hline													
LINMIX\_ERR (median posterior distribution)	&	1.25	$\pm$	0.22	&	1.21	$\pm$	0.40	&	0.11	$\pm$	0.05	\\
\hline
\end{tabular}
\end{minipage}
\end{table*}



\begin{thebibliography}{99}

\bibitem[Adelman-McCarthy
et al.(2007)]{2007ApJS..172..634A} Adelman-McCarthy, J.~K., et al.\ 2007, ApJS, 172, 634

\bibitem[Akritas
\& Bershady(1996)]{1996ApJ...470..706A} Akritas, M.~G., \& Bershady, M.~A.\ 1996, ApJ, 470, 706

\bibitem[Bentz
et al.(2009)]{2009ApJ...705..199B} Bentz, M.~C., et al.\ 2009, \apj, 705, 199

\bibitem[Collin
et al.(2006)]{2006AAp...456...75C} Collin, S., Kawaguchi, T., Peterson, B.~M., \& Vestergaard, M.\ 2006, A\&A, 456, 75

\bibitem[Denney
et al.(2006)]{2006ApJ...653..152D} Denney, K.~D., et. al.\ 2006, \apj, 653, 152

\bibitem[Denney
et al.(2009)]{2009ApJ...702.1353D} Denney, K.~D., et. al.\ 2009, \apj, 702, 1353

\bibitem[Denney
et al.(2010)]{2010ApJ...721..715D} Denney, K.~D., et. al.\ 2010, \apj, 721, 715

\bibitem[Grier et al.(2008)]{2008ApJ...688..837G} Grier, C.~J., et. al.\ 2008, \apj, 688, 837

\bibitem[Haario et al.
(2006)]{2006..............H} Haario, H., Laine, M., Mira, A., \& Saksman, E., {\em Dram: Efficient Adaptive MCMC}, Statistics and Computing, Springer Netherlands 16, no. 4, 339-354

\bibitem[Kelly,
B. C.(2007)]{2007ApJ...665.1489K} Kelly, B.~C.\ 2007, \apj, 664, 1489

\bibitem[Kollmeier et
al.(2006)]{2006ApJ...648..128K} Kollmeier J.~A., et al., 2006, ApJ, 648, 128

\bibitem[McLure
\& Jarvis(2002)]{2002MNRAS.337..109M} McLure, R.~J., \& Jarvis, M.~J.\ 2002, MNRAS, 337, 109

\bibitem[McLure
\& Dunlop(2004)]{2004MNRAS.352.1390M} McLure, R.~J., \& Dunlop, J.~S.\ 2004, MNRAS, 352, 1390

\bibitem[Metzroth
et. al.(2006)]{2006ApJ...647..901M} Metzroth, K.~G., Onken, C.~A., \& Peterson, B.~M.\ 2006, \apj, 647, 901

\bibitem[Netzer
\& Trakhtenbrot(2007)]{2007ApJ...654..754N} Netzer, H., \& Trakhtenbrot, B.\ 2007, ApJ, 654, 754

\bibitem[Onken
\& Kollmeier(2008)]{2008ApJ...689L..13O} Onken, C.~A., \& Kollmeier, J.~A.\ 2008, ApJL, 689, L13

\bibitem[Peterson et al.
(2004)]{2004ApJ...613..682P} Peterson, B.~M., et al.\ 2004, \apj, 613, 682

\bibitem[Peterson,
B. M.(2007)]{2007ASPC..373....3P} Peterson, B.~M.\ 2007, in The Central Engine of Active Galactic Nuclei, ed.  Luis C. Ho \& Jian-Min Wang, ASP Conference Series, Vol. 373, 3

\bibitem[Rafiee
\& Hall(2011)]{2011ApJS..........R} Rafiee, A., \& Hall, P.~B.\ 2011, ApJS, submitted

\bibitem[Schneider
et al.(2007)]{2007AJ....134..102S} Schneider, D.~P., et al.\ 2007, AJ, 134, 102

\bibitem[Shen et al.(2008)]{2008ApJ...680..169S} Shen, Y., Greene, J.~E.,
Strauss, M.~A., Richards, G.~T., \& Schneider, D.~P.\ 2008, ApJ, 680, 169

\bibitem[Shen et al.(2010)]{2010arXiv1006.5178S} Shen, Y., et al.\ 2010, submitted (arXiv:1006.5178)

\bibitem[Steinhardt
\& Elvis (2010a)]{2010MNRAS.402.2637S} Steinhardt, C.~L., \& Elvis, M.\ 2010, MNRAS, 402, 2637 (SE10a)

\bibitem[Steinhardt
\& Elvis (2010b)]{2010MNRASb} Steinhardt, C.~L., \& Elvis, M.\ 2010, MNRAS, 410, 201 

\bibitem[Steinhardt
\& Elvis (2010c)]{2010MNRAS.406L...1S} Steinhardt, C.~L., \& Elvis, M.\ 2010, MNRAS, 406, L1

\bibitem[Steinhardt
\& Elvis (2010d)]{2010d} Steinhardt, C.~L., \& Elvis, M.\ 2010, arXiv:1011.6381

\bibitem[Sulentic et al.(2000)]{2000ApJ...536L...5S} Sulentic, J.~W.,
Zwitter, T., Marziani, P., \& Dultzin-Hacyan, D.\ 2000, \apjl, 536, L5

\bibitem[Tremaine et al.(2002)]{2002ApJ...574..740T} Tremaine, S., et al.\
2002, ApJ, 574, 740

\bibitem[Vestergaard
\& Peterson(2006)]{2006ApJ...641..689V} Vestergaard, M., \& Peterson, B.~M.\ 2006, ApJ, 641, 689

\bibitem[Wang et al.(2009)]{2009ApJ...707.1334W} Wang, J.-G., et al.\ 2009,
\apj, 707, 1334

\end{thebibliography}
\end{document}